\documentclass[preprint,preprintnumbers,amsmath,amssymb]{revtex4}

\usepackage{graphicx}
\usepackage{dcolumn}
\usepackage{bm}

\begin{document}

\preprint{Proceeding "International Conference on Nuclear Data for Science and Technology 2007"}

\title{New Stellar $(n,\gamma)$ Cross Sections and The "Karlsruhe Astrophysical Database of Nucleosynthesis in Stars"}

\author{I. Dillmann}\email{iris.dillmann@ik.fzk.de}
 \altaffiliation[also at ]{Departement f\"ur Physik und Astronomie, Universit\"at Basel.}
\author{R. Plag}\altaffiliation[now at ]{Gesellschaft f\"ur Schwerionenforschung mbH, Plankstrasse 1, D-64291 Darmstadt}
\author{C. Domingo-Pardo}
\author{F. K\"appeler}
 \affiliation{Institut f\"ur Kernphysik, Forschungszentrum Karlsruhe, Postfach 3640, D-76021 Karlsruhe,
        Germany}
\author{M. Heil}
   \affiliation{Gesellschaft f\"ur Schwerionenforschung mbH, Plankstrasse 1, D-64291 Darmstadt }
        
\author{T. Rauscher}
\author{F.-K. Thielemann}
 \affiliation{Departement Physik und Astronomie, Universit\"at Basel, Klingelbergstrasse 82, CH-4056 Basel,
        Switzerland}

\begin{abstract}
Since April 2005 a regularly updated stellar neutron cross section compilation is available online at http://nuclear-astrophysics.fzk.de/kadonis. This online-database is called the "Karlsruhe Astrophysical Database of Nucleosynthesis in Stars" project and is based on the previous Bao et al. compilation from the year 2000. The present version \textsc{KADoNiS} v0.2 (January 2007) includes recommended cross sections for 280 isotopes between $^{1}$H and $^{210}$Po and 75 semi-empirical estimates for isotopes without experimental information. Concerning stellar $(n,\gamma)$ cross sections of the 32 stable, proton-rich isotopes produced by the $p$ process experimental information is only available for 20 isotopes, but 9 of them have rather large uncertainties of $\geq$9\%. The first part of a systematic study of stellar $(n,\gamma)$ cross sections of the $p$-process isotopes $^{74}$Se, $^{84}$Sr, $^{102}$Pd, $^{120}$Te, $^{130}$Ba, $^{132}$Ba, $^{156}$Dy, and $^{174}$Hf is presented. In another application \textsc{KADoNiS} v0.2 was used for an modification of a reaction library of Basel university. With this modified library $p$-process network calculations were carried out and compared to previous results.
\end{abstract}

\maketitle
\section{Stellar neutron capture compilations} \label{intro}
The pioneering work for stellar neutron capture cross sections was published in 1971 by Allen and co-workers \cite{alle71}. In this
paper the role of neutron capture reactions in the
nucleosynthesis of heavy elements was reviewed and a list of
recommended (experimental or semi-empirical) Maxwellian averaged
cross sections at $kT$= 30 keV (MACS30) presented for nuclei between
C and Pu.

The idea of an experimental and theoretical stellar neutron cross
section database was picked up again by Bao and K\"appeler
\cite{bao87} for $s$-process studies. This compilation
published in 1987 included cross sections for ($n,\gamma$)
reactions (between $^{12}$C and $^{209}$Bi), some ($n,p$) and
($n,\alpha$) reactions (for $^{33}$Se to $^{59}$Ni), and
also ($n,\gamma$) and ($n,f$) reactions for
long-lived actinides. A follow-up compilation was published by
Beer et al. in 1992 \cite{BVW92}.

In the update of 2000 the Bao compilation \cite{bao00} was extended
down to $^{1}$H and -- like the original Allen paper -- semi-empirical
re\-commended values for nuclides without experimental cross
section information were added. These estimated values are normalized cross
sections derived with the Hauser-Feshbach code NON-SMOKER
\cite{rath00}, which account for known systematic deficiencies in
the nuclear input of the calculation. Additionally, the database
provided stellar enhancement factors and energy-dependent MACS for
energies between $kT$= 5 keV and 100 keV.

The \textsc{KADoNiS} project \cite{kado06} is based on these previous compilations and aims to be a regularly updated database. The current version \textsc{KADoNiS} v0.2 (January 2007) is already the second update and includes - compared to the previous Bao et al. compilation \cite{bao00}- 38 updated and 14 new recommended cross sections. The update history can be followed in the section ``Logbook''. A paper version of \textsc{KADoNiS} (``v1.0'') is planned for 2008, which also will -- like the first Bao compilation from 1987 \cite{bao87}-- include ($n,p$) and ($n,\alpha$) reactions for light isotopes and ($n,\gamma$) and ($n,f$) reactions for long-lived actinides at $kT$= 30~keV. Additionally, a re-calculation of semi-empirical estimates based on the latest experimental results of neighboring nuclides will be performed.

\section{Systematic study of $(n,\gamma)$ cross sections for the $p$ process}
\subsection{The ``$p$ processes''}
The ``$p$ process'' is responsible for the production of 32 stable but rare isotopes between $^{74}$Se and $^{196}$Hg on the proton-rich side of the valley of stability. Unlike the remaining 99\% of the heavy nuclei beyond iron these isotopes cannot be created by neutron captures in the $s$ process or $r$ process, and their solar \cite{AnG89} and isotopic abundances \cite{iupac} are 1-2 orders of magnitude lower than the respective $s$- and $r$-process nuclei. The bulk of $p$ isotopes is thought to be produced in explosive O/Ne burning during supernova type II explosions (core collapse supernovae). This mechanism is called ``$\gamma$ process'' since the main reactions are photo-induced reactions of high energy photons ($T_9$= 2-3) on pre-existing seed nuclei from prior $s$-processing. The ``$\gamma$ process'' can reproduce the solar abundances \cite{AnG89} of most $p$ isotopes within a factor of 3 \cite{ray90,raar95}. 

For the missing abundances of the most abundant isotopes $^{92,94}$Mo and $^{96,98}$Ru alternative processes have been proposed, e.g. using strong neutrino fluxes in the ``$\nu p$ process'' \cite{FML06}, or rapid proton-captures in the ``$rp$ process'' \cite{scha98} in a binary, cataclysmic system with a neutron star accreting material from a Red Giant.

Apart from the astrophysical uncertainties of the $p$-process site another problem arises from the large nuclear physics uncertainties due to missing experimental cross section data. Present network calculations for the reproduction of the solar abundances of the 32 $p$-process isotopes include up to now only the $\gamma$-process scenario since it seems to be the best understood part of the $p$ processes. Such network calculations are carried out with a typical reaction library consisting of $\approx$1600 isotopes which are connected by several thousands of reactions \cite{RGW06,ID06}. The largest fraction of these reactions concerns short-lived radioactive nuclei and thus have to be inferred from theoretical work, e.g. the Hauser-Feshbach statistical model \cite{hafe52,rath00}.

Some experimental information for the $p$ process is available for charged-particle reactions, but the largest amount of data concerns $(n,\gamma)$ data, which is connected via detailed balance with the respective $(\gamma,n)$ reactions needed for the $\gamma$ process. However, most of this neutron capture data was measured with the activation technique at one single energy ($kT$=25~keV), and thus has to be extrapolated to the respective $\gamma$-process energies ($kT$=170-260~keV) with the help of energy-dependencies from Hauser-Feshbach theory. This method was used in the modification of a reaction library discussed in Sec.~\ref{sim}. 

\subsection{Experimental technique}
The previous status of $(n,\gamma)$ cross sections for the 32 $p$ isotopes at $kT$= 30
keV is listed in the third column of Table \ref{tab:4} \cite{bao00}. Experimental data was available for 20 isotopes but 9 of them ($^{92,94}$Mo, $^{96}$Ru, $^{124,126}$Xe, $^{130}$Ba, $^{156}$Dy,
$^{180}$W, and $^{190}$Pt) exhibited uncertainties larger than 9\%. For the remaining 12 isotopes no experimental information was available in the stellar energy range, where only semi-empirical estimates based on Hauser-Feshbach predictions existed.

This motivated an extended measuring campaign at the Karlsruhe 3.7 MV Van de Graaff
accelerator using the activation technique. In this framework the $p$ isotopes $^{74}$Se, $^{84}$Sr, $^{102}$Pd, $^{120}$Te, $^{130}$Ba, $^{132}$Ba, $^{156}$Dy, and $^{174}$Hf were measured with samples of natural composition. Neutrons were produced via the $^{7}$Li$(p,n)$$^{7}$Be reaction by bombarding 10-30 $\mu$m thick layers of metallic lithium or lithiumfluoride
on a water-cooled copper backing with protons of E$_p$= 1912 keV, 31 keV above the $^{7}$Li($p,n$) reaction threshold at 1881~keV. The resulting quasi-stellar neutron spectrum approximates a Maxwellian distribution for $kT$= \mbox{25.0 $\pm$ 0.5 keV} \cite{raty88} but is truncated at $E_n$= 106~keV. Under these conditions, all neutrons are kinematically collimated into a forward cone of 120$^\circ$ opening angle. Neutron scattering through the Cu backing is negligible, since the transmission is $\approx$98\% in
the energy range of interest. 

The sample materials were either metals (Se, Pd, Te, Dy, and Hf) or compounds (SrO, SrCO$_{3}$, SrF$_{2}$, BaCO$_{3}$). Thin pellets were pressed from the respective powders or granules
and enclosed in cans made from thin aluminium foil. In case of Pd, Dy, and Hf the samples were cut from thin metal foils. During the activations the samples were sandwiched between 10-30 $\mu$m thick gold foils of the same diameter and were irradiated in close geometry to the neutron target. In this way the neutron flux can be determined relative to the well-known capture cross section of $^{197}$Au \cite{raty88}. The activation measurements were carried out with the Van de Graaff accelerator operated in DC mode with a current of $\approx$100~$\mu$A (for the Li targets) or even higher currents (up to 150~$\mu$A) for the LiF targets. The mean neutron flux over the period of the activations was $\approx$1.5-3$\times$10$^9$ s$^{-1}$ at the position of the samples. To ensure homogeneous illumination of the entire surface the proton beam was wobbled across the Li target. During the irradiation the neutron flux was recorded in intervals of 60~s or 90~s using a $^6$Li-glass detector for later correction of the number of nuclei which decayed during the activation.

For the measurement of the induced activities two detector setups were available. A single high purity Germanium (HPGe) detector with a well defined geometry and 10~cm
lead shielding was used in all cases for the counting of the gold foils, as well as for the activities of $^{75}$Se, $^{85}$Sr, $^{121}$Te, $^{131}$Ba, $^{133m}$Ba, $^{157}$Dy, and $^{175}$Hf. The activities of $^{103}$Pd and $^{133g}$Ba were measured with a gamma detection system consisting of two HPGe Clover detectors \cite{DPA04} in close geometry. The decay properties of the determined product nuclei are given in Table~\ref{tab:1}. The sample and activation parameters are shortly summarized in Table \ref{tab:2}. 

\begin{table}[!h]
\caption[Decay properties]{\label{tab:1}Decay properties of
the product nuclei. Shown here are only the strongest transitions used for the analysis.}
\begin{center}
\renewcommand{\arraystretch}{1.2} 
\begin{tabular}{ccccc}
Isotope 		& t$_{1/2}$ & E$_\gamma$ [keV] & I$_\gamma$ [\%] & Ref.\\
\hline
$^{75}$Se   & 119.79 (4)~d  & 136.0 & 58.3 (7) & \cite{nds75} \\
            &               & 264.7 & 58.9 (3) & \\
\hline
$^{85}$Sr$\rm ^g$ & 64.84 (2)~d    	& 514.0 & 95.7 (40) & \cite{nds85}\\
$^{85}$Sr$\rm ^m$ & 67.63 (4)~min 	& 151.2 & 12.9 (7)  & \\
            			&             		& 231.9 & 84.4 (22)  & \\
\hline
$^{103}$Pd  & 16.991 (19)~d 		& 357.5 & 2.21$\times$10$^{-2}$ (7) & \cite{nds103}\\
\hline
$^{121}$Te$\rm ^g$ & 19.16 (5)~d    & 573.1 & 80.3 (25) & \cite{nds121}\\
$^{121}$Te$\rm ^m$ & 154 (7)~d  		& 212.2 & 81.4 (1)  & \\
               			&            		& 1102.1 & 2.54 (6) & \\
\hline
$^{131}$Ba$\rm ^g$ & 11.50 (6)~d   & 123.8 & 29.0 (3) & \cite{nds131} \\
               &               & 216.1 & 19.7 (2) & \\
               &               & 373.2 & 14.0 (2) & \\
               &               & 496.3 & 46.8 (2) & \\
\hline
$^{133}$Ba$\rm ^g$ & 10.52 (13)~yr 	& 356.0 & 62.1 (2) & \cite{nds133a} \\
$^{133}$Ba$\rm ^m$ & 38.9 (1)~h  		& 275.9 & 17.8 (6) & \cite{nds133b} \\
\hline
$^{157}$Dy     & 8.14 (4)~h  		& 326.3 & 92 (4) & \cite{nds157} \\
\hline
$^{175}$Hf     & 70 (2)~d      	& 343.4 & 84.0 (30) & \cite{nds175} \\
\hline
$^{198}$Au     & 2.69517 (21)~d & 411.8 & 95.58 (12) & \cite{nds198} \\
\hline
\end{tabular}
\end{center}
\end{table}

\begin{table}[!htb]
\caption{\label{tab:2}Activation schemes and sample characteristics. $\Phi$$_{tot}$ is the total neutron
exposure of the sample during the activation.}
\begin{center}
\renewcommand{\arraystretch}{1.2} 
\begin{tabular}{cccc}
\hline
Isotope & N(Isotope) & t$_{act}$  & $\Phi_{tot}$ \\
				& [atoms]	   & [min]		& [neutrons] \\
\hline
$^{74}$Se & (0.7-1.4)$\times$10$^{19}$ & 419-1425 & (0.2-1.5)$\times$10$^{14}$ \\
$^{84}$Sr$\rightarrow$$\rm ^g$ & (0.2-1.3)$\times$10$^{19}$ & 1234-2621 & (0.8-1.6)$\times$10$^{14}$ \\
$^{84}$Sr$\rightarrow$$\rm ^m$ & (0.4-0.8)$\times$10$^{19}$ & 155-274   & (1.1-3.3)$\times$10$^{13}$ \\
$^{102}$Pd & (1.7-2.6)$\times$10$^{19}$ & 5751-9770 & (3.5-8.2)$\times$10$^{14}$ \\
$^{120}$Te$\rightarrow$$\rm ^{g,m}$ & (1.6-2.0)$\times$10$^{18}$ & 1406-4142 & (1.5-3.1)$\times$10$^{14}$\\
$^{130}$Ba & (3.5-4.8)$\times$10$^{17}$ & 4014-7721 & (2.7-6.9)$\times$10$^{14}$ \\
$^{132}$Ba$\rightarrow$$\rm ^{g,m}$ & (3.3-4.7)$\times$10$^{17}$ & 4014-7721 & (2.7-6.9)$\times$10$^{14}$\\
$^{156}$Dy & (0.6-1.8)$\times$10$^{17}$ & 362-964 & (4.0-9.5)$\times$10$^{13}$ \\
$^{174}$Hf & (5.0-8.8)$\times$10$^{17}$ & 3865-5451 & (5.1-8.7)$\times$10$^{14}$ \\
\hline
\end{tabular}
\end{center}
\end{table}

\subsection{Results}
For a detailed description of the data analysis and the results, see \cite{ID06,DHK06,VDK06}. The resulting Maxwellian averaged cross sections at $kT$=30~keV from this measuring campaign are shown in Table~\ref{tab:4} in bold. The semi-empirical estimates for $^{74}$Se, $^{102}$Pd, $^{120}$Te, $^{132}$Ba, and $^{174}$Hf are reproduced within the large error bars of the prediction. The previous experimental values for $^{130}$Ba and $^{156}$Dy are confirmed perfectly but with much improved uncertainties. 

Thus, only 6 $p$ isotopes ($^{98}$Ru, $^{138}$La, $^{158}$Dy, $^{168}$Yb, $^{184}$Os, and $^{196}$Hg) remain without any experimental information about the stellar neutron cross section. The present work is therefore being extended to the heavier $p$ isotopes and includes the measurement of $^{158}$Dy, $^{168}$Yb, $^{184}$Os, and $^{196}$Hg, and the re-measurement of $^{180}$W and $^{190}$Pt. Only for $^{98}$Ru and $^{138}$La the activation technique cannot be applied.

\begin{table}[!htb]
\centering
\caption{Status of MACS30 of all 32 $p$ nuclei. Isotopic abundances were taken from \cite{iupac}. Recommended cross section were taken from \cite{bao00} and \cite{kado06}. \emph{Italic values} show semi-empirical estimates. *Preliminary semi-empirical value based on \textsc{KADoNiS} v0.2. **Preliminary value, data analysis not yet fully completed.} \label{tab:4}
\renewcommand{\arraystretch}{1.2} 
\begin{tabular}{cccc}
   \hline\noalign{\smallskip}
Isotope & Isotopic & \multicolumn{2}{c}{Recommended MACS30} \\
				& abundance \cite{iupac} & previous \cite{bao00} & new \cite{kado06} \\
				& [\%] & [mb] & [mb] \\
\hline
 $^{74}$Se & 0.89 (4) & \emph{267 $\pm$ 25} & \textbf{271 $\pm$ 15} \cite{DHK06}\\
 $^{78}$Kr & 0.35 (1) & \multicolumn{2}{c}{312 $\pm$ 26} \\
 $^{84}$Sr & 0.56 (1) & \emph{368 $\pm$ 125} & \textbf{300 $\pm$ 17} \cite{DHK06}\\
 $^{92}$Mo & 14.84 (35) & \multicolumn{2}{c}{70 $\pm$ 10} \\
 $^{94}$Mo & 9.25 (12) & \multicolumn{2}{c}{102 $\pm$ 20} \\
 $^{96}$Ru & 5.54 (14) & 238 $\pm$ 60 & 207 $\pm$ 8 \\
 $^{98}$Ru & 1.87 (3) & \multicolumn{2}{c}{\emph{173 $\pm$ 36}} \\
 $^{102}$Pd & 1.02 (1) & \emph{373 $\pm$ 118} & \textbf{370 $\pm$ 14 }\\
 $^{106}$Cd & 1.25 (6) & \multicolumn{2}{c}{302 $\pm$ 24} \\
 $^{108}$Cd & 0.89 (3) & \multicolumn{2}{c}{202 $\pm$ 9} \\
 $^{113}$In & 4.29 (5) & \multicolumn{2}{c}{787 $\pm$ 70} \\
 $^{112}$Sn & 0.97 (1) & \multicolumn{2}{c}{210 $\pm$ 12} \\
 $^{114}$Sn & 0.66 (1) & \multicolumn{2}{c}{134.4 $\pm$ 1.8}\\
 $^{115}$Sn & 0.34 (1) & \multicolumn{2}{c}{342.4 $\pm$ 8.7} \\
 $^{120}$Te & 0.09 (1) & \emph{420 $\pm$ 103} & \textbf{499 $\pm$ 24}  \\
 $^{124}$Xe & 0.09 (1) & \multicolumn{2}{c}{644 $\pm$ 83} \\
 $^{126}$Xe & 0.09 (1) & \multicolumn{2}{c}{359 $\pm$ 51} \\
 $^{130}$Ba & 0.106 (1) & 760 $\pm$ 110 & \textbf{767 $\pm$ 30} \\
 $^{132}$Ba & 0.101 (1) & \emph{379 $\pm$ 137} & \textbf{399 $\pm$ 16} \\
 $^{136}$Ce & 0.185 (2) & \multicolumn{2}{c}{328 $\pm$ 21}  \\
 $^{138}$Ce & 0.251 (2) & \multicolumn{2}{c}{179 $\pm$ 5} \\
 $^{138}$La & 0.090 (1) & \multicolumn{2}{c}{\emph{419 $\pm$ 59*}} \\
 $^{144}$Sm & 3.07 (7) & \multicolumn{2}{c}{92 $\pm$ 6} \\
 $^{156}$Dy & 0.06 (1) & 1567 $\pm$ 145 & \textbf{1548 $\pm$ 30**} \\
 $^{158}$Dy & 0.10 (1) & \multicolumn{2}{c}{\emph{1060 $\pm$ 400}} \\
 $^{162}$Er & 0.14 (1) & \multicolumn{2}{c}{1624 $\pm$ 124} \\
 $^{168}$Yb & 0.13 (1) & \multicolumn{2}{c}{\emph{1160 $\pm$ 400}} \\
 $^{174}$Hf & 0.16 (1) & \emph{956 $\pm$ 283}$^*$ & \textbf{983 $\pm$ 46}  \cite{VDK06}\\
 $^{180}$W  & 0.12 (1) & \multicolumn{2}{c}{536 $\pm$ 60} \\
 $^{184}$Os & 0.02 (1) & \multicolumn{2}{c}{\emph{657 $\pm$ 202}} \\
 $^{190}$Pt & 0.014 (1)& \multicolumn{2}{c}{677 $\pm$ 82} \\
 $^{196}$Hg & 0.15 (1) & \multicolumn{2}{c}{\emph{650 $\pm$ 82}} \\
\hline
\end{tabular}
\end{table}

\section{$p$-process simulations with an updated reaction library}\label{sim}
The $p$-process network calculations in \cite{RGW06,ID06} were carried out with the program "\textsc{pProSim}" \cite{RGW06,rapp04}. The underlying network was originally based on a reaction library from Michigan State University for X-ray bursts, which included only proton-rich isotopes up to Xenon. For $p$-process studies it was extended with a full reaction network library from Basel university \cite{reaclib}. This reaction library is mainly based on NON-SMOKER predictions \cite{rath00} with only a few experimental information for light nuclei, and was modified with the latest stellar neutron capture cross sections available from \textsc{KADoNiS} v0.2. This modification includes more than 350 experimental and semi-empirical ($n,\gamma$) cross sections and was extended to the respective ($\gamma$,n) channels calculated via detailed balance. 

\textsc{pProSim} simulates the abundance evolution for the 32 $p$ isotopes with a parameterized model of a
supernova type II explosion of a 25 M$_\odot$ star. Since the $p$-process
layers are located far outside the collapsing core, they only experience the bounced shock front passing through the O/Ne burning zone and the subsequent temperature and density increase. Both, the seed abundances and the respective temperature and density profiles, were taken from external works and not calculated self-consistently (for more information, see \cite{RGW06}).

The results from the simulations with the modified reaction library were compared to the results published in \cite{RGW06} to examine the influence of the experimental neutron capture data. This was done with help of the so-called ``normalized overproduction factor'', which is =1 when the calculated abundance corresponds to the solar abundances \cite{AnG89}. Ranges of variations of this factor for SN type II explosions with star masses 13~$M_\odot$$\leq M_\star$$\leq$25~$M_\odot$ are published e.g. in Fig.~4 in \cite{raar95}.

The new overproduction factors are slightly below previously published values \cite{raar95,RGW06} due to the inclusion of recent experimental data, especially in the mass range 150$\leq$$A$$\leq$170 where the main reaction flux is driven by $(\gamma,n)$, $(\gamma,\alpha)$, $(n,\gamma)$, and $(n,\alpha)$ reactions. All of these reaction fluxes are found to be smaller. Our study underlines the importance of $(n,\gamma)$ and $(\gamma,n)$ reactions in the $p$-process flow. For example, we were able to show that a variation in the neutron rates of the Pb and Bi isotopes has a strong impact on the above mentioned fluxes.
This is due to the fact that a significant fraction of the seed abundances is located in these isotopes and converted to nuclei at lower mass by photodisintegration sequences starting with $(\gamma,n)$ reactions on Pb and Bi. Also the importance of experimental data is strongly emphasized by these findings. Because of the magicity or near-magicity of the Pb and Bi isotopes, individual resonances determine the cross sections and the Hauser-Feshbach theory is not applicable \cite{LL60,RTK96}. From discrepancies between resonance and activation measurements \cite{MHW77,BCM97} and from theoretical considerations \cite{RBO98}, it has been previously found that even a small direct capture component contributes to neutron capture on Pb \cite{LL60,RBO98}. Resonant and direct capture contributions are difficult to handle in theoretical models and experiments prove to be indispensable.

\begin{acknowledgments}
We thank E. P. Knaetsch, D. Roller and W. Seith for their help and
support during the neutron irradiations at the Karlsruhe Van de Graaff accelerator.
This work was supported by the Swiss National Science Foundation
Grants 2024-067428.01 and 2000-105328.
\end{acknowledgments}

\end{document}